\documentclass[showpacs,floatfix,eqsecnum,nofootinbib,aps]{revtex4}

\usepackage{amssymb}
\usepackage{latexsym}
\usepackage{epsfig}
\usepackage{citesort}

\def\ds{\displaystyle}
\def\beq{\begin{equation}}
\def\eeq{\end{equation}}
\def\bea{\begin{eqnarray}}
\def\eea{\end{eqnarray}}
\def\beeq{\begin{eqnarray}}
\def\eeeq{\end{eqnarray}}
\def\la{\langle}
\def\ra{\rangle}
\def\gggg{\gamma \gamma \to \gamma \gamma}

\newcommand{\cleqn}{\setcounter{equation}{0}}

\newcommand{\iden}{\leavevmode\hbox{\small1\normalsize\kern-.33em1}}

\begin{document}




\title{Little Higgs model effects at $\gamma \gamma$ collider}

\author{S. Rai Choudhury}
   \email{src@physics.du.ac.in}
   \affiliation{Department of Physics $\&$ Astrophysics,
     University of Delhi, Delhi - 110 007, India.}
\author{A. S. Cornell}
   \email{alanc@yukawa.kyoto-u.ac.jp}
   \affiliation{Yukawa Institute for Theoretical Physics,
    Kyoto University, Kyoto 606-8502, Japan.}
\author{Naveen Gaur 
\footnote{Present address : Theory Group, KEK, Oho 1-1, 
Tsukuba 305-0801, Japan }}
   \email{naveen@post.kek.jp, naveen@physics.du.ac.in}
   \affiliation{Department of Physics $\&$ Astrophysics,
     University of Delhi, Delhi - 110 007, India.}
\author{Ashok Goyal}
    \email{agoyal@iucaa.ernet.in}
   \affiliation{Department of Physics $\&$ Astrophysics,
     University of Delhi, Delhi - 110 007, India.}

\begin{abstract}
Though the predictions of the Standard Model (SM) are in
excellent agreement with experiments there are still several
theoretical problems, such as fine-tuning and the hierarchy
problem. These problems are associated with the Higgs sector of the
SM, where it is widely believed that some {\it ``new physics"} will
take over at the TeV scale. One beyond the SM theory which resolves
these problems is the Little Higgs (LH) model. In this work we shall
investigate the effects of the LH model on $\gggg$ scattering; where
the process $\gggg$ at high energies occurs in the SM through diagrams
involving $W$, charged quark and lepton loops (and is, therefore,
particularly sensitive to any new physics). 
\end{abstract}

\date{\today}

\maketitle

%
%

\section{Introduction}\cleqn

\par It has been known for some time that the $\gggg$ scattering
amplitude at high energies will be a very useful tool in the search
for new particles and interactions in an $e^+ e^-$ linear collider
operated in the $\gamma \gamma$ mode. In particular, as according to
present ideas, this scattering can be achieved by colliding $e^{\pm}$
beams at a future linear collider, such as the ILC, with laser photons
(which are subsequently backscattered, through the Compton effect) to
produce very energetic photons of high luminosity along the
$e^{\pm}$ direction; while the $e^{\pm}$ beams would loose most of
their energy. As such, these searches may involve either the direct
production of new degrees of freedom (for example, charginos, light
sleptons or a light stop in SUSY models); or the precise study of the
production of SM particles, where the the new degrees of freedom
contribute virtually in some loop diagrams. In this respect, processes
like $\gggg$, $\gamma \gamma \to Z \gamma$, $\gamma \gamma \to Z Z$
should all provide very important tools for searching or constraining
new physics \cite{Choudhury:1999gp}; particularly as the SM
contributions in these processes first appear at the one-loop level
and should be small.    

\par As a large number of helicity amplitudes can contribute to these
processes, due to the presence of spin-one particles in the initial
and final states, considerations of symmetries and other invariances
is required to reduce this number. Furthermore, in the SM the
amplitudes of $\gggg$ will have contributions from one-loop diagrams
mediated by charged fermions (leptons and quarks) and $W$-bosons. At
large energies ($\sqrt{s_{\gamma \gamma}} \ge 250 GeV$) it is know
that the $W$ contributions dominate over the fermionic
contributions. At these energies it should also be noted that the
dominant amplitudes are predominantly imaginary. Therefore we expect
that any new physics effects in the $\gggg$ process may come from the
interference terms between the predominantly imaginary SM amplitudes
and new physics effects to these amplitudes.  

\par At this point we would like to point out that though the SM has
been very successful in explaining all electroweak interactions probed
so far, there is no symmetry or relation which protects the mass of
the Higgs boson. In fact the Higgs mass diverges quadratically when
quantum corrections in the SM are taken into account. But precision
electroweak data demands the lightest Higgs boson mass be $\sim
200$GeV! In order for this to happen we either need to invoke some
symmetries which will protect the Higgs mass to a much higher scale
(possibly GUT scale) or assume that the SM is an effective theory
valid up to only the electroweak scale. In either of these
possibilities it is expected that some new physics should takeover
from the SM at the TeV scale. As such, Supersymmetry has provided one
popular example of new physics, where additional symmetries are
invoked which help protect the Higgs mass up to GUT scale. Recently a
new approach to address this problem has been advocated, the approach
popularly known as the {\it ``Little Higgs 
models''}, which addresses some of the problems in the SM by making
the Higgs boson a  pseudo-Goldstone boson of a symmetry which is
broken at some higher scale $\Lambda$. The suggestion of making the
Higgs a pseudo-Goldstone boson was proposed some time ago
\cite{Georgi:1975tz} but has been revived recently, where such models
have been successfully constructed by Arkani-Hamed, Cohen and Georgi
\cite{Arkani-Hamed:2001nc}. The successful Little Higgs models are
constructed in such a way that {\it no single} interaction breaks all
the symmetries, but the symmetries are broken {\it collectively}. In
these models the scale $\Lambda$ ($= 4 \pi f$) is chosen to be $\sim
10$TeV. The scale $\Lambda$ acts as a cut-off which separates the
weakly interacting low energy range from possible strongly interacting
sectors at higher energies. The Higgs fields then acquire a mass
radiatively at the electroweak scale. Note 
that in this model the Higgs field remains light, being protected by
the approximate global symmetry and free from any one-loop quadratic
sensitivity to the cutoff scale $\Lambda$. Note also, that in doing
this we are required to introduce several new heavy gauge bosons and 
other new particles, which shall be discussed further in section 2. 

\par However, it must be noted that the originally proposed
implementations of the LH approach suffered from severe constraints
from precision electroweak measurements \cite{Yue:2004xt}, which could
only be satisfied by finely tuning the model parameters. The most
serious constraint resulted from the tree-level corrections to
precision electroweak observables due to the exchanges of the
additional heavy gauge bosons present in the theories (because their
masses are much smaller than the cut-off scale), as well as from the
small but non-vanishing vev of an additional weak-triplet scalar
field. As a result, masses of new particles had to be raised, and the
fine-tuning of the Higgs boson mass is re-introduced. Motivated by
these constraints, several new variants of the LH model were proposed
\cite{Chang:2003un}. Particularly interesting is the 
implementation of the $Z_2$ symmetry, called T-parity, into the model,
as proposed in references \cite{Cheng:2005as}. T-parity explicitly
forbids any tree-level contribution from the new heavy gauge bosons to
the observables involving only SM particles as external states. It
also 
forbids the interactions that induced the triplet vev. As a result, in
T-parity symmetric LH models, corrections to precision electroweak
observables are generated exclusively at loop level. This implies that
the constraints are generically weaker than in the tree-level case,
and fine tuning can be avoided
\cite{Hubisz:2005tx}.    

\par Note that due to T-parity the lightest T-odd particle becomes
stable and a good candidate for dark matter. This is an
interesting feature of the model, because the existence of dark matter
is now established by recent cosmological observations
\cite{Asano:2006nr}. Since the lightest T-odd particle is
electrically and colour neutral, and has a mass of ${\cal O}(100)$GeV
\cite{Cheng:2005as,Asano:2006nr} in many LH models with T-parity, 
these models provide a WIMP dark matter candidate \cite{Cheng:2005as},
and are able to account for the large scale structure of the present
universe.  

\par With this in mind, we review the LH model we have used in section 2
before proceeding to investigate the helicity amplitudes of the
scattering process $\gggg$ in section 3. Finally we conclude with the
discussion of the results of our numerical analysis in section 4.  

%
%

\section{Little Higgs models}\cleqn

\par In this section we will briefly describe the LH models
which we have used in our analysis. In particular the minimal version
of the LH model, the so-called {\sl Littlest Higgs model}
\cite{Han:2003wu}.   

\par To begin, let us recall that it is known that the scalar mass in
a generic quantum field theory will receive quadratically divergent
radiative corrections, all the way up to the cut-off scale. The LH
model solves this problem by eliminating the lowest 
order contributions via the presence of a partially broken global
symmetry (where the non-linear transformation of the Higgs fields
under this global symmetry prohibits the existence of a Higgs mass
term of the form $m^2 |h|^2$). This is done by introducing a new set
of heavy gauge bosons (with the same quantum numbers as the SM gauge
bosons), where the gauge couplings to the Higgs bosons are patterned
in such a way that the quadratic divergences induced in the SM gauge
boson loops are canceled by the quadratic divergence induced by the
heavy gauge bosons at one-loop level. One also introduces a heavy
fermionic state which couples to the Higgs field in a specific way, so
that the one-loop quadratic divergence induced by the top-quark
Yukawa coupling to the Higgs boson is canceled. Furthermore, extra
Higgs fields exist as the Goldstone boson multiplets from the global
symmetry breaking. On this framework the {\sl Littlest Higgs} model
was introduced, which was based on an $SU(5)/SO(5)$ coset. The
phenomenology of this model was discussed in great detail in precision
tests \cite{Han:2003wu,Yue:2004xt} and low energy measurements
\cite{Buras:2004kq}. LH models generically also predict the existence
of a doubly charged triplet Higgs. The phenomenology of triplet Higgs
within the context of the LH model has also been extensively studied
in the literature \cite{Han:2005nk}. A detailed review of LH models
can be found in \cite{Schmaltz:2005ky}. 

The model we shall use in our analysis, the {\sl Littlest Higgs
model} \cite{Han:2003wu,Schmaltz:2005ky}, is a non-linear $\sigma$
model based on an 
$SU(5)$ global symmetry which contains a gauged $\left[SU(2) 
\times U(1) \right]_1\bigotimes \left[SU(2) \times U(1) \right]_2$
symmetry with couplings $g_1, g_2, g_1'$ and $g_2'$ as its
subgroup. Furthermore, the global $SU(5)$ symmetry is broken into
$SO(5)$ by the vacuum expectation value of the sigma field   
\beq
\Sigma_0 = \left( \begin{array}{ccc}
0 & 0 & \iden \\
0 & 1 & 0 \\
\iden & 0 & 0 
\end{array} \right) .
\label{eq:sec2:1}
\eeq
Where $\iden$ is the $2\times 2$ identity matrix. This breaking
simultaneously breaks the gauge group to an $SU(2) \times U(1)$
subgroup, which is identified with the SM group. The breaking of the
global $SU(5) \to SO(5)$ gives rise to 14 goldstone bosons, $\Pi =
\pi^a X^a$ which can be written as 
\beq
\Sigma = \ds e^{i \Pi/f} \Sigma_0 e^{i \Pi^T/f} 
= \Sigma_0 + \frac{2 i}{f} \Pi \Sigma_0 + {\cal O}(1/f^2) ,
\label{eq:sec2:2}
\eeq
where $X^a$ corresponds to the broken $SU(5)$ generators. Four of the
fourteen Goldstone bosons are absorbed by the broken gauge generators,
and the remaining ten Goldstones are parameterized as:  
\beq
\Pi = \left(
\begin{array}{ccc}
            & h^\dagger/\sqrt{2}  &  \Phi^\dagger   \\
h/\sqrt{2}  &   & h^*/\sqrt{2}                \\
\Phi        & h^T/\sqrt{2}  &    
\end{array}
\right) ,
\label{eq:sec2:3}
\eeq
where $h$ is the SM Higgs doublet and $\Phi$ is a complex $SU(2)$
triplet\footnote{the existence of the $SU(2)$ triplet is a generic
feature of LH models}:  
\beq
\Phi = \left(
\begin{array}{cc}
\Phi^{++} & \Phi^+/\sqrt{2}     \\
\Phi^+/\sqrt{2}  & \Phi^0
\end{array}
\right) .
\label{eq:sec2:4}
\eeq
The kinetic term for the $\Sigma$ field can be written as 
\beq 
{\cal L}_{kin} = \ds \frac{f^2}{8} \mathrm{Tr} \left\{D_{\mu} \Sigma (
D^{\mu} \Sigma)^{\dagger} \right\}, \label{eq:sec2:5} 
\eeq
where
\beq
D_{\mu} \Sigma = \partial_{\mu} \Sigma - i \Sigma_j \left[ g_j W_j^a
(Q_j^a \Sigma + \Sigma Q_j^{aT}) + g'_j B_j (Y_j \Sigma + \Sigma Y_j )
\right] .  
\label{eq:sec2:6}
\eeq
In the above $j = 1, 2$, the $Q_j$ and $Y_j$ are the gauged
generators, $B_j$ and $W_j^a$ are the $U(1)_j$ and $SU(2)_j$ gauge
fields, respectively, and $g_j$ and $g'_j$ are the corresponding
coupling constants.   

\par As stated earlier, the vev ($\Sigma_0$) given in
eqn(\ref{eq:sec2:1}) breaks the gauge group to the diagonal one, which
is then identified with the SM group. This generates mass and mixings
of the gauge bosons. The heavy gauge boson mass eigenstates are given
by   
\beq
W_H^a = - c W_1^a + s W_2^a ~~,~~ B_H = - c' B_1 + s' B_2 ,
\label{eq:sec2:7} 
\eeq
where $s, s', c$ and $c'$ are the mixing angles given by
\bea
c' =& g'/g_2' ~~~,~~~ s' &= g'/g_1' , \nonumber \\
c =& g/g_2 ~~~,~~~ s &= g/g_1 .
\label{eq:sec2:8}
\eea
These couplings can be related to the SM couplings ($g, g'$) by
\cite{Han:2003wu}:
\beq
\frac{1}{g^2} = \frac{1}{g_1^2}  +\frac{1}{g_2^2} ~~~,~~~
\frac{1}{g'^2} = \frac{1}{g_1'^2}  +\frac{1}{g_2'^2} ,
\label{eq:sec2:9}
\eeq
where the masses of heavy gauge bosons will then be:
\beq
M_{W_H}^2 = \frac{f^2}{4} \left(g_1^2 + g_2^2\right) ~~,~~
M_{B_H}^2 = \frac{f^2}{20} \left(g_1'^2 + g_2'^2\right) .
\label{eq:sec2:10}
\eeq
The orthogonal combination of these gauge bosons are identified with
the SM $W$ and $B$.    

\par In the SM the top quark introduces quadratic corrections to the
Higgs boson mass. The LH model addresses this problem by the
introduction of a new set of heavy fermions which couple to the Higgs
such that it cancels the quadratic divergences to the Higgs mass; due
to the SM top quark. A vector like top quark is usually introduced in
the LH model to do this job. The Yukawa interactions in the LH model are
chosen to be:  
\beq
{\cal L}_Y = {1 \over 2} \lambda_1 f \epsilon_{ijk} \epsilon_{xy}
\chi_i \Sigma_{jx} \Sigma_{ky} u_3'^c  + \lambda_2 f \tilde{t}
\tilde{t}'^c + h.c. \label{eq:sec2:11} 
\eeq
with $i,j,k$ summed over 1, 2, 3 and $x$, $y$ summed over 4,
5. $\chi_i = (b_3,t_3,\tilde{t})$, $b_3$ and $t_3$ are the SM bottom
and top quarks, $(\tilde{t},\tilde{t}'^c)$ is the new vector like top
quark and $u_3'^c$ is the SM right handed top quark.   

\par Expanding the $\Sigma$ field and diagonalizing the mass matrix we
arrive at the physical states:  
\beq
m_t \sim  \frac{v \lambda_1 \lambda_2}{\sqrt{\lambda_1^2 +
\lambda_2^2}} ~~~~,~~~~ 
m_T \sim  f \sqrt{\lambda_1^2 + \lambda_2^2} . \label{eq:sec2:12}
\eeq
These masses are parameterized in term of $x_L$, defined as:
\beq
x_L = \frac{\lambda_1^2}{\lambda_1^2 + \lambda_2^2} . \label{eq:sec2:13}
\eeq
In the LH models there is no Higgs potential at tree level, this is
generated at one-loop level via the interactions with gauge bosons and
fermions. This is similar to a Coleman-Weinberg type of
potential. This gives the Higgs masses as:  
\beq
M_\Phi^2 = 2 m_H^2\frac{f^2}{v^2} \frac{1}{1 - \left(\frac{4 v'
f}{v^2}\right) } , \label{eq:sec2:14} 
\eeq
where $m_H$ is the SM Higgs boson mass. Therefore, for this kind of LH
model (based on $SU(5)/SO(5)$) we have five input parameters, in
addition to the SM Higgs mass, explicitly, these are   
$$
s, s', x_L, f, v' .
$$

\par The advantage of this model now becomes apparent; by noting that
as the gauge generators are embedded in the $SU(5)$ group, in such a
way as to commute with an $SU(3)$ subgroup, one pair of gauge
couplings must be set to zero. Therefore the Higgs mass would be an
exact Goldstone boson and massless. As such, any diagram renormalizing
the Higgs mass will vanish unless it involves at least two of the
gauge couplings. Note that at the one-loop level all diagrams
satisfying this condition are only logarithmically
divergent. Therefore the symmetry breaking mechanism protects the
Higgs mass from quadratic divergences at this level. Generically the
particle spectrum of the LH model, apart from SM particles is:  
\begin{itemize}
\item{} Heavy vector like top quark (T);
\item{} Heavy gauge bosons: charged ($W_H$), neutral( $Z_H, A_H$);
\item{} Additional triplet Higgs: ($\Phi^0, \Phi^+ , \Phi^{++}$).
\end{itemize}

\par As mentioned in the introduction, the original LH models were
severely constrained by precision electroweak experiments
\cite{Han:2003wu,Yue:2004xt}. The main constraints coming from the
$\rho$ parameter and the $Z \to b \bar{b}$ \cite{Yue:2004xt} vertex
contributions. Other models which evade these constraints have been
proposed, but all of these enlarge the global or gauge
symmetries. Recently Cheng and Low \cite{Cheng:2005as}
introduced a discrete 
$Z_2$ symmetry, which we now call {\it ``T-parity"} to resolve the electroweak
precision constraints in the LH models. The advantages of introducing
T-parity is two fold. Firstly it helps relaxing the precision
constraints and secondly it also provides a dark matter
candidate. The new parity is an exchange of the two gauge
groups and the Lagrangian in eqn (\ref{eq:sec2:5})
is invariant under this exchange provided that $g_1 = g_2$ and
$g'_1 = g'_2$. The implication of this is that the gauge boson mass
eigenstates will have the form $W_{\pm} = \frac{1}{\sqrt{2}}
(W_1 \pm W_2)$ and $B_{\pm} = \frac{1}{\sqrt{2}}(B_1 \pm B_2)$. The SM
gauge bosons are even under T-parity and are designated by a $+$
subscript, and the new {\it ``T-odd"} gauge bosons are designated by a
$-$ subscript. The different T-parity states do not mix and
after electroweak symmetry breaking, the Weinberg angle is given by the
usual SM relation, as are other electroweak observables (thus removing
the constraint). Note further that as the transformation law ensures
that the complex $SU(2)$ triplet is odd under T-parity, whilst the
Higgs doublet is even, the trilinear coupling $H^{\dagger} \phi H$ is
forbidden. This further relaxes precision electroweak constraints on
the model.   

\par Thus the main implications of the introduction of T-parity are: 
\begin{itemize}
\item{} All new particles (except one heavy top quark) are odd under
T-parity;  
\item{} T-parity exchanges $[SU(2)\times U(1)]_1$ and $[SU(2)\times
U(1)]_2$;  
\item{} T-parity imposes a relationship between the couplings, for
example $g_1 = g_2$, $g_1' = g_2'$;  
\item{} The fermion sector is extended to include T-odd fermions; 
\item{} There is no vev to the triplet Higgs. This being assured by
the absence of a $H\Phi H$ coupling.   
\end{itemize}
In such a model ($SU(5)/SO(5)$ with T-parity) the input model
parameters (apart from SM Higgs mass $m_H$) are: 
$$
f~,~~ \frac{\lambda_1}{\lambda_2}, ~~ \kappa .
$$
The first two have been defined before and $\kappa$ is a free
parameter whose range is $0.5 \le \kappa \le 1.5$
\cite{Hubisz:2005tx}. As we are interested in the $\gggg$ process,
where this process occurs at one-loop by the  
mediation of charged particles, it should be noted that in the SM this
process is mediated by charged $W$ and fermions (charged leptons and
quarks). This process in the LH model can also be mediated by charged
gauge bosons ($W_H$), charged Higgs ($\Phi^-, \Phi^{--}$) and new
fermions ($T$ in the LH model without T-parity; note that in the LH
model with T-parity, there shall also be $T_+, T_-$ and heavy T-odd
fermions which can mediate the process). To conclude this section, we
have given the mass spectrum of these particles in the LH models in
Table \ref{table:1}.  

\begin{table}
\begin{center}
\begin{tabular}{|c | c  c  c  c  c  c |} \hline 
Particle &  $W_H$ & $\Phi$ & $m_t$ & $T_-$ & $T_+$ & $f_H$ \\  \hline
 & & & & & & \\ 
Masses (LH with T-parity)  &  $g f$  & $\frac{\sqrt{2} m_H f}{v}$ &  
$\frac{\lambda_1 \lambda_2 v}{\sqrt{\lambda_1^2 + \lambda_2^2}}$ & 
$\lambda_2 f$    & $\sqrt{\lambda_1^2 + \lambda_2^2} f$ & $\kappa f$ 
 \\  \hline
 & & & & & &  \\ 
Masses (LH without T-parity)  &
$\frac{g f}{2 s c}$ & 
$\frac{\sqrt{2} m_H f/v}{\sqrt{\left[1 - (4 v'f/v^2)^2 \right]}}$ & 
$\frac{\lambda_1 \lambda_2 v}{\sqrt{\lambda_1^2 + \lambda_2^2}}$ & - 
&  - & - 
\\ \hline
\end{tabular}
\end{center}
\caption{\it Mass spectrum of the LH models. $v$ is the vev of the SM
Higgs and $m_H$ is the SM Higgs mass. In the model without T-parity we
have only one vector like top quark with mass $m_T = \sqrt{\lambda_1^2
+ \lambda_2^2} f$. $f_H$ are the T-odd fermions in the LH with
T-parity.}   
\label{table:1}
\end{table}

%
%

\section{The $\gamma \gamma \to \gamma \gamma$ crossections}\cleqn

\par The process
\beq
\gamma (p_1, \lambda_1) \gamma (p_2, \lambda_2) \to \gamma (p_3,
\lambda_3) \gamma (p_4, \lambda_4)      
\eeq
can be represented by sixteen possible helicity amplitudes
$F_{\lambda_1 \lambda_2 \lambda_3 \lambda_4}(\hat{s}, \hat{t},
\hat{u})$, where the $p_i$ and $\lambda_i$ represent the respective 
momenta and helicities; the $\hat{s}$, $\hat{t}$ and $\hat{u}$ are the
usual Mandelstam variables. By the use of Bose statistics, crossing
symmetries and demanding parity and time-invariance, these sixteen
possible helicity amplitudes can be expressed in terms of just three 
amplitudes, namely (the relationship between various helicity
amplitudes as given in appendix \ref{appen:a})  
\beq
F_{++++}(\hat{s}, \hat{t}, \hat{u}) , \hspace{0.5cm} F_{+++-}(\hat{s},
\hat{t}, \hat{u}) , \hspace{0.5cm} F_{++--}(\hat{s}, \hat{t}, \hat{u})
.   
\eeq
As such, the cross-section for this process can be expressed as
\cite{Jikia:1993tc}:
\bea
\ds \frac{d \sigma}{d\tau d\cos\theta^*} & = & \ds
\frac{d\bar{L}_{\gamma\gamma}}{d\tau} \left\{ \frac{d
\bar{\sigma}_0}{d\cos\theta^*} + \la \xi_2 \xi'_2 \ra \frac{d
\bar{\sigma}_{22}}{d\cos\theta^*} + \left[ \la \xi_3 \ra \cos 2 \phi +
\la \xi'_3 \ra \cos 2 \phi \right] \right. \nonumber \\  
& & \hspace{0.5cm} \times \ds \frac{d \bar{\sigma}_3}{d\cos\theta^*} +
\la \xi_3 \xi'_3 \ra \left[ \frac{d \bar{\sigma}_{33}}{d\cos\theta^*}
\cos 2 ( \phi + \phi' ) + \frac{d \bar{\sigma}'_{33}}{d \cos\theta^*}
\cos 2 ( \phi - \phi' ) \right] \nonumber \\  
& & \left. \hspace{0.5cm} \ds + \left[ \la \xi_2 \xi'_3 \ra \sin 2
\phi' - \la \xi_3 \xi'_2 \ra \sin 2 \phi' \right] \frac{d
\bar{\sigma}_{23}}{d \cos\theta^*} \right\} ,  
\eea
where $d \bar{L}_{\gamma \gamma}$ describes the photon-photon
luminosity in the $\gamma \gamma$ mode and $\tau = s_{\gamma
\gamma}/s_{e e}$. Note that $\xi_2$, $\xi'_2$, $\xi_3$ and $\xi'_3$ 
are the Stokes parameters. Furthermore
\cite{Jikia:1993tc,Gounaris:1999gh},   
\bea
\frac{d \bar{\sigma}_0}{d\cos\theta^*} & = & \left( \frac{1}{128 \pi
\hat{s}} \right) \sum_{\lambda_3 \lambda_4} \left[ \left|
F_{++\lambda_3 \lambda_4} \right|^2 + \left| F_{+-\lambda_3 \lambda_4}
\right|^2 \right] , \nonumber \\  
\frac{d \bar{\sigma}_{22}}{d\cos\theta^*} & = & \left( \frac{1}{128
\pi \hat{s}} \right) \sum_{\lambda_3 \lambda_4} \left[ \left|
F_{++\lambda_3 \lambda_4} \right|^2 - \left| F_{+-\lambda_3 \lambda_4}
\right|^2 \right] , \nonumber \\  
\frac{d \bar{\sigma}_3}{d\cos\theta^*} & = & \left( -\frac{1}{64 \pi
\hat{s}} \right) \sum_{\lambda_3 \lambda_4} \mathrm{Re} \left[
F_{++\lambda_3 \lambda_4} F^*_{-+\lambda_3 \lambda_4} \right] ,
\nonumber \\  
\frac{d \bar{\sigma}_{33}}{d\cos\theta^*} & = & \left( \frac{1}{128
\pi \hat{s}} \right) \sum_{\lambda_3 \lambda_4} \mathrm{Re} \left[
F_{+-\lambda_3 \lambda_4} F^*_{-+\lambda_3 \lambda_4} \right] ,
\nonumber \\  
\frac{d \bar{\sigma}'_{33}}{d\cos\theta^*} & = & \left( \frac{1}{128
\pi \hat{s}} \right) \sum_{\lambda_3 \lambda_4} \mathrm{Re} \left[
F_{++\lambda_3 \lambda_4} F^*_{--\lambda_3 \lambda_4} \right] ,
\nonumber \\  
\frac{d \bar{\sigma}_{23}}{d\cos\theta^*} & = & \left( \frac{1}{64 \pi
\hat{s}} \right) \sum_{\lambda_3 \lambda_4} \mathrm{Im} \left[
F_{++\lambda_3 \lambda_4} F^*_{+-\lambda_3 \lambda_4} \right] .  
\eea
To obtain the total cross-section from the above expressions the
integration over $cos\theta^*$ has to be done in the range $0 \le
\cos\theta^* \le 1$. However, the whole range of $\theta^*$ will not be
experimentally observable, hence, for our numerical estimates we will
restrict the scattering angle to be $|\cos\theta^*| \le \cos30^\circ$. 
It should be noted that of the above mentioned cross-sections only $d
\bar{\sigma}_0 / d\cos\theta^*$ should be positive, where the angle
$\theta^*$ is the scattering angle of the photons in $\gamma \gamma$
rest frame. The process $\gggg$ proceeds through the mediation of
charged particles. In the SM these charged particles were charged
gauge bosons ($W$), quarks and charged leptons. In the LH model, in
addition to the charged gauge bosons and fermions, we also have
charged scalars. The analytical expressions of the contributions from
fermions, gauge bosons and scalars to the helicity amplitudes are
given in \cite{Jikia:1993tc} and are quoted in  
Appendix \ref{appen:a}.  With these equations in hand we shall, in the
next section, analyse what effects the LH models will have on these
cross-sections.   

%
%

\section{Results and Conclusions}\cleqn

\par In this section we shall present our numerical analysis of the
$\gggg$ scattering in the LH model, with and without T-parity. Note
that in the $\gamma \gamma$ scattering process the helicity amplitudes
are proportional to the fourth power of the charge of the particle
circulating in the box {\sl i.e.}  
$$ F_{\lambda_1 \lambda_2 \lambda_3 \lambda_4} \propto Q^4 , $$ 
where $Q$ is the charge of the particle. In the LH models we
generically have a triplet of scalar particles one of which is doubly
charged, such as $\Phi^{--}$. This results in a factor of 16 in the
amplitude and hence a factor of 256 in the cross-section. This should
provide noticeable signatures in the cross-sections.   

\par
\begin{figure}
\begin{center}
\epsfig{file=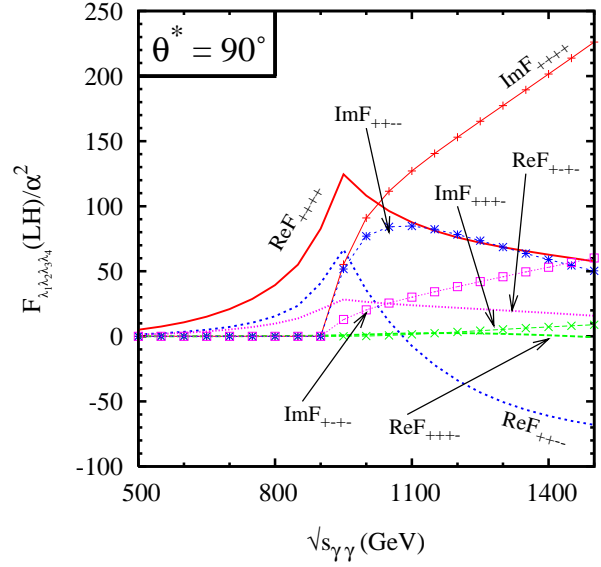,width=.65\textwidth}
\end{center}
\vskip -.5cm
\caption{\it Helicity amplitudes of the LH (with T-parity)
contributions for $\theta^* = 90^\circ$. For this plot we have taken
$f = 700$ GeV. For $\theta^* = 90^\circ$ we have $F_{+--+} = F_{+-+-}$.}
\label{fig:lh_amp_1}
\end{figure}

\begin{figure}
\hskip -1.5cm
\epsfig{file=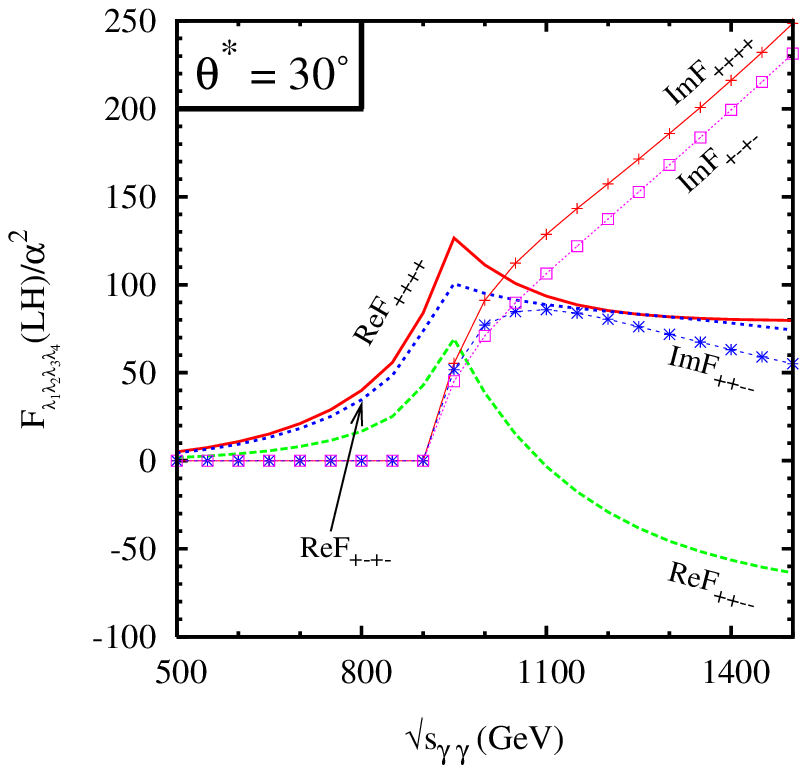,width=.6\textwidth} \hskip -3cm
\epsfig{file=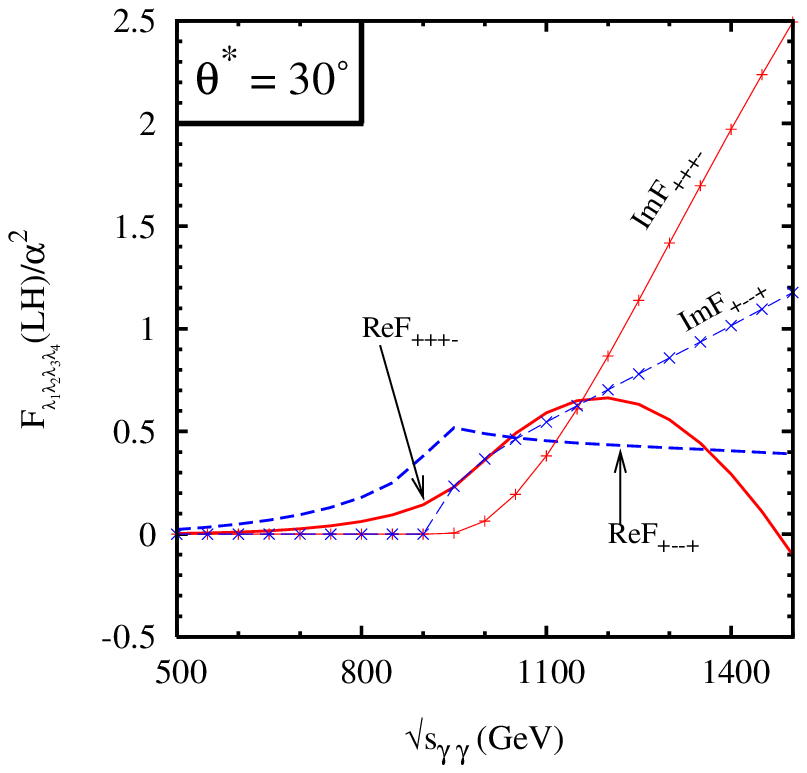,width=.6\textwidth}
\vskip -.5cm
\caption{\it Helicity amplitudes of LH (with T-parity) contributions
for $\theta^* = 30^\circ$.}   
\label{fig:lh_amp_2}
\end{figure}

 In our first set of results, presented in figures
(\ref{fig:lh_amp_1},\ref{fig:lh_amp_2}), we have shown the
contribution of LH particles to the various helicity amplitudes
introduced earlier. In figure (\ref{fig:lh_amp_1}) we have shown the
behavior of both real and imaginary parts of the helicity amplitudes
for $\theta^* = 90^\circ$ and in figure (\ref{fig:lh_amp_2}) the
results have been plotted for 
$\theta^* = 30^\circ$. Note that for a scattering angle ($\theta^*$) of
$90^\circ$ we have $\hat{u} = \hat{t}$, which results in $F_{+--+} =
F_{+-+-}$. Whereas this relationship is not present for other values
of the scattering angle. 

\par
\begin{figure}
\vspace*{-.5cm}
\hspace*{-1cm}
\epsfig{file=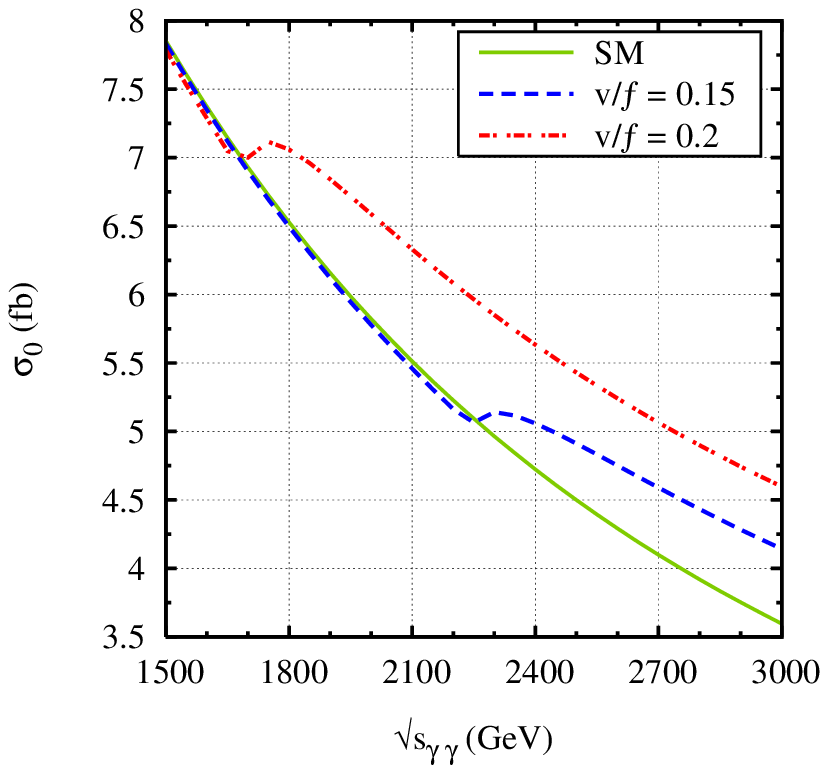,width=.6\textwidth} \hskip -3 cm
\epsfig{file=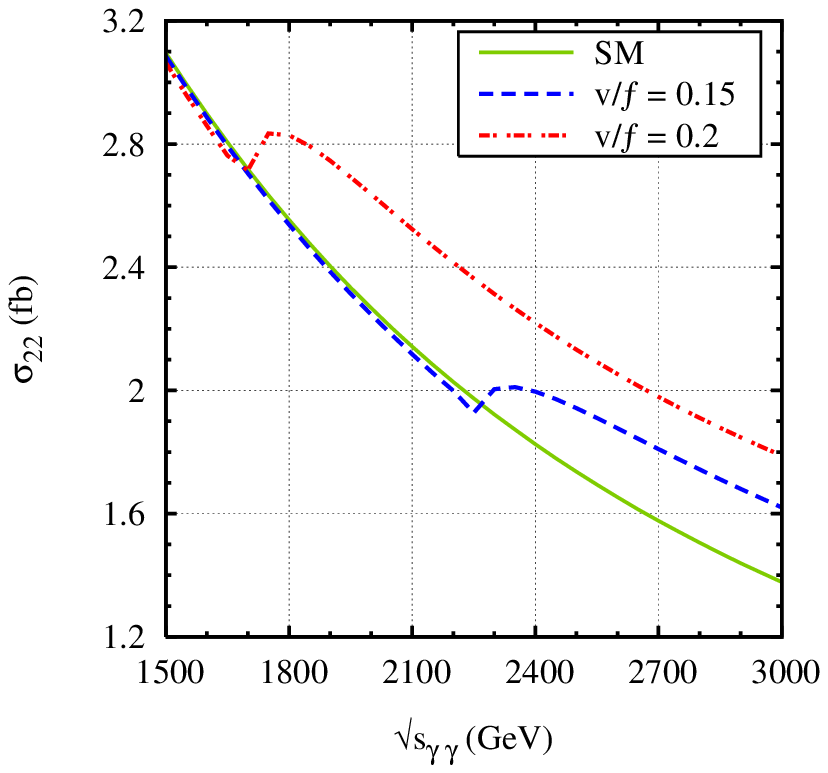,width=.6\textwidth} \\
\vspace*{-.2cm}
\hspace*{-1cm}
\epsfig{file=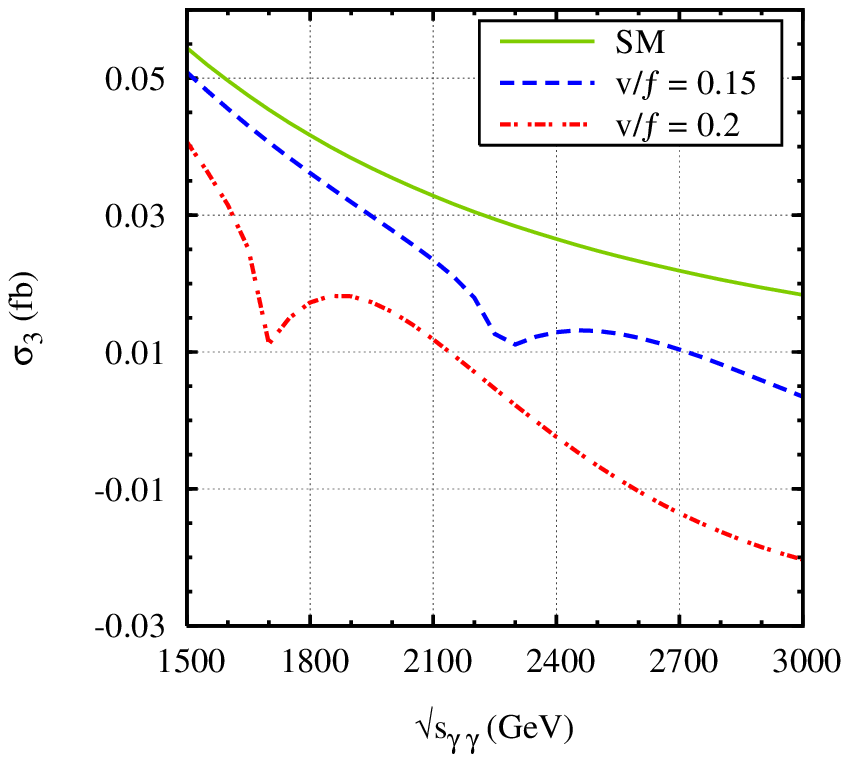,width=.6\textwidth} \hskip -3 cm
\epsfig{file=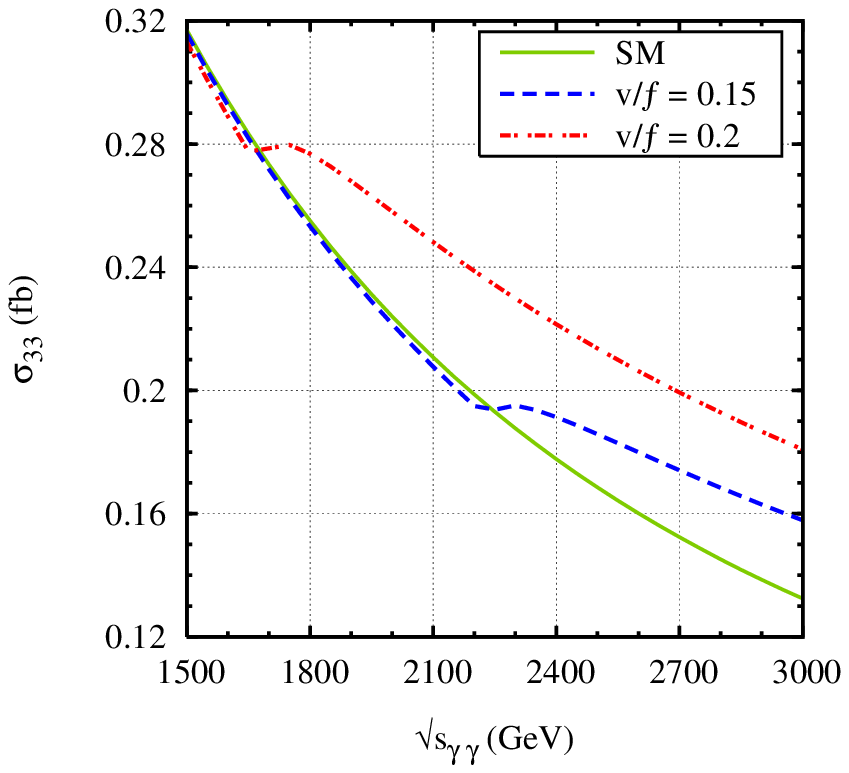,width=.6\textwidth}
\vspace*{-.2cm}
\hspace*{-1cm}
\epsfig{file=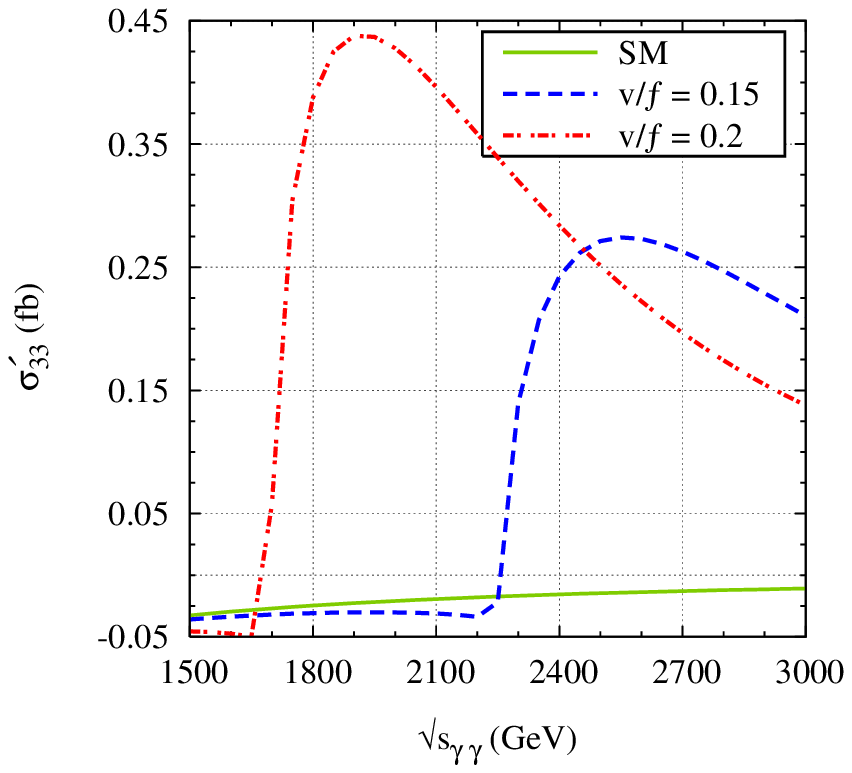,width=.6\textwidth} \hskip -3 cm
\epsfig{file=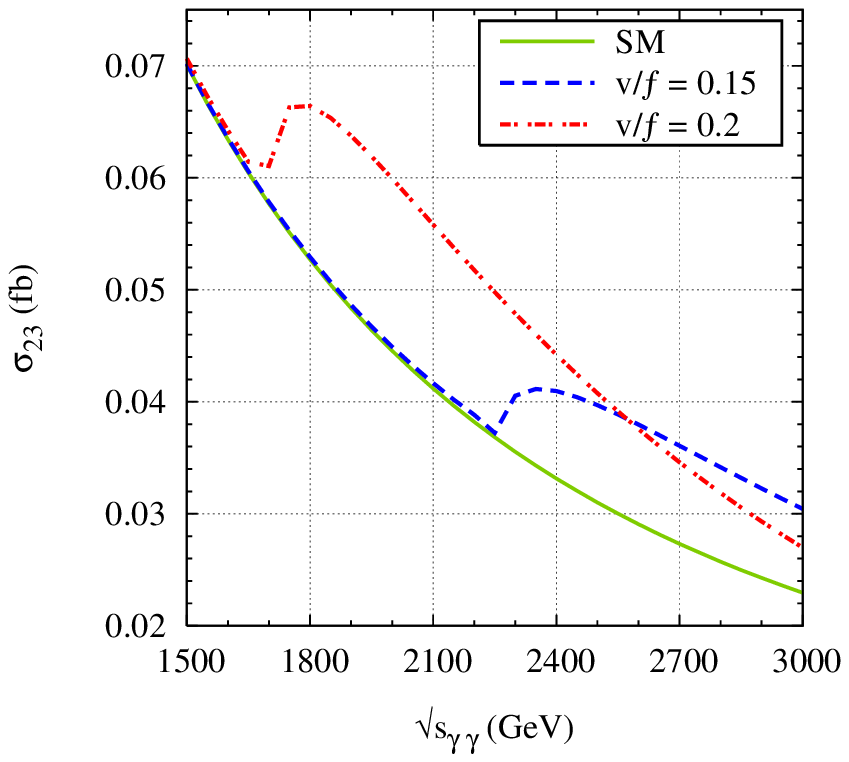,width=.6\textwidth}
\vskip -.5cm
\caption{\it Results for the cross-sections integrated in the range
$30 \le \theta^* \le 150$ for various values of $v/f$. Other LH model
parameters are: $x_L = 0.2, s = s' = 0.6$.}  
\label{fig:lh_intcross_1}
\end{figure}

\begin{figure}
\hspace*{-1cm}
\epsfig{file=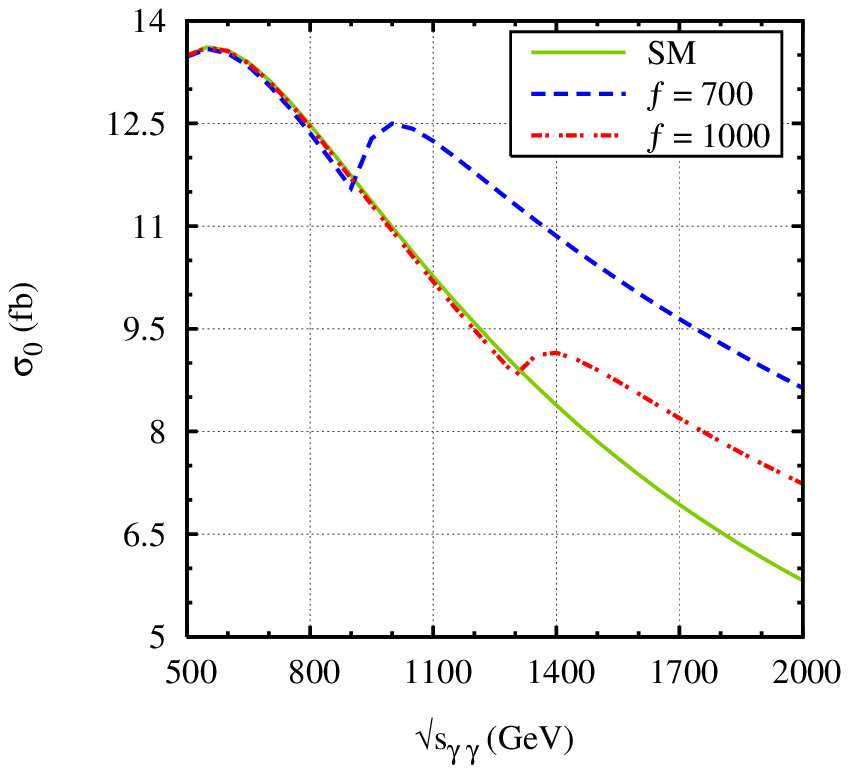,width=.6\textwidth} \hskip -3 cm
\epsfig{file=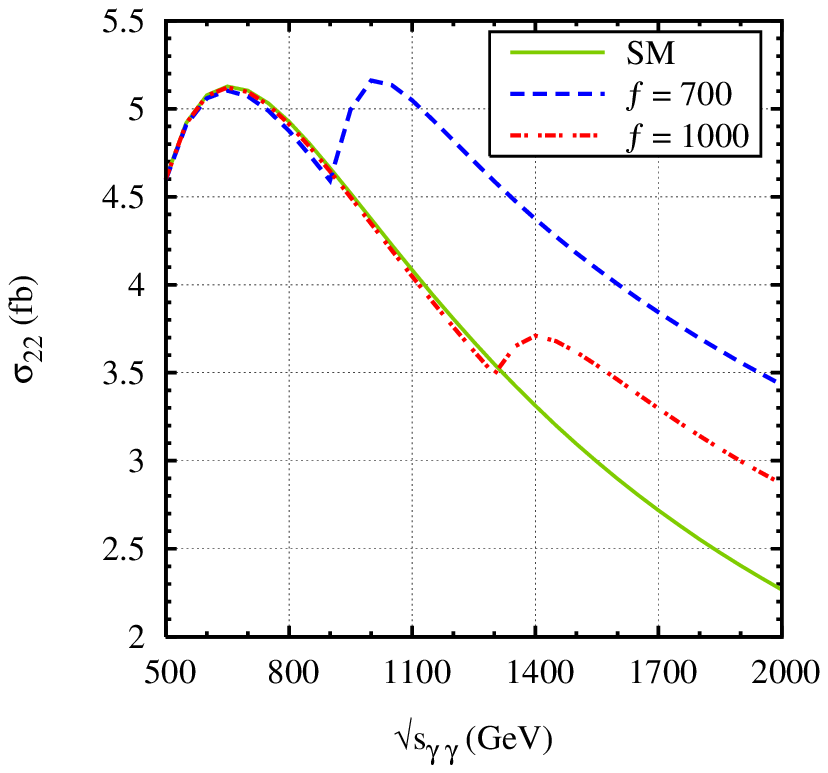,width=.6\textwidth}
\vspace*{-.2cm}
\hspace*{-1cm}
\epsfig{file=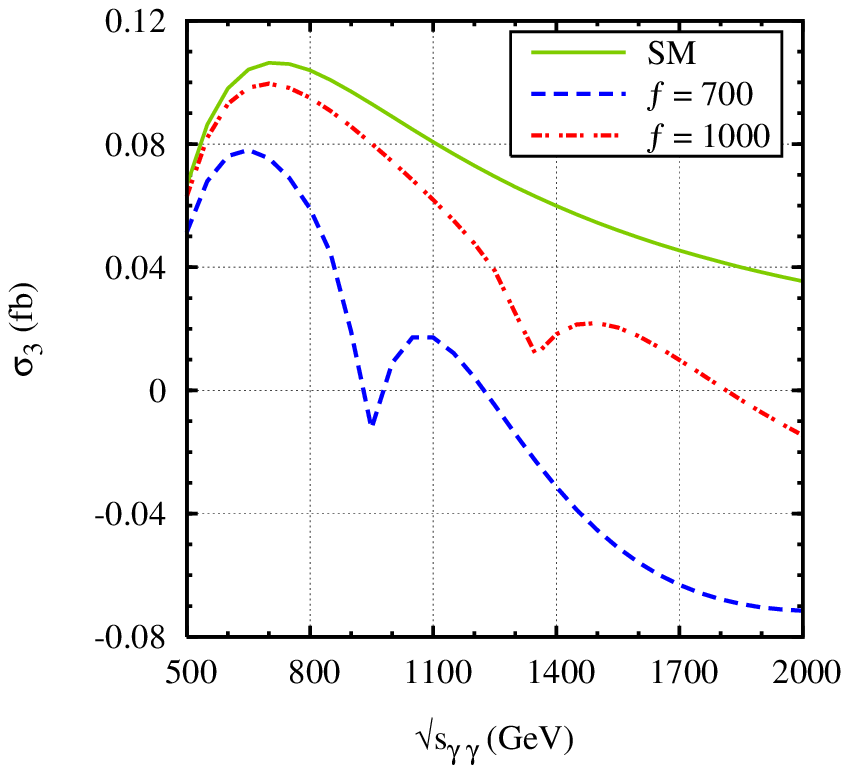,width=.6\textwidth} \hskip -3 cm
\epsfig{file=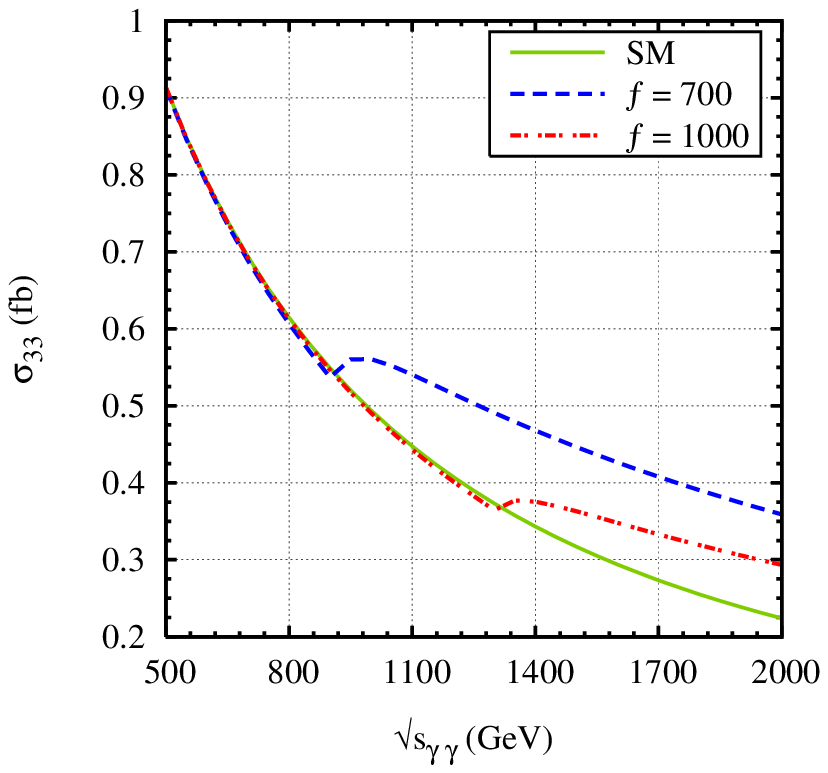,width=.6\textwidth}
\vspace*{-.2cm}
\hspace*{-1cm}
\epsfig{file=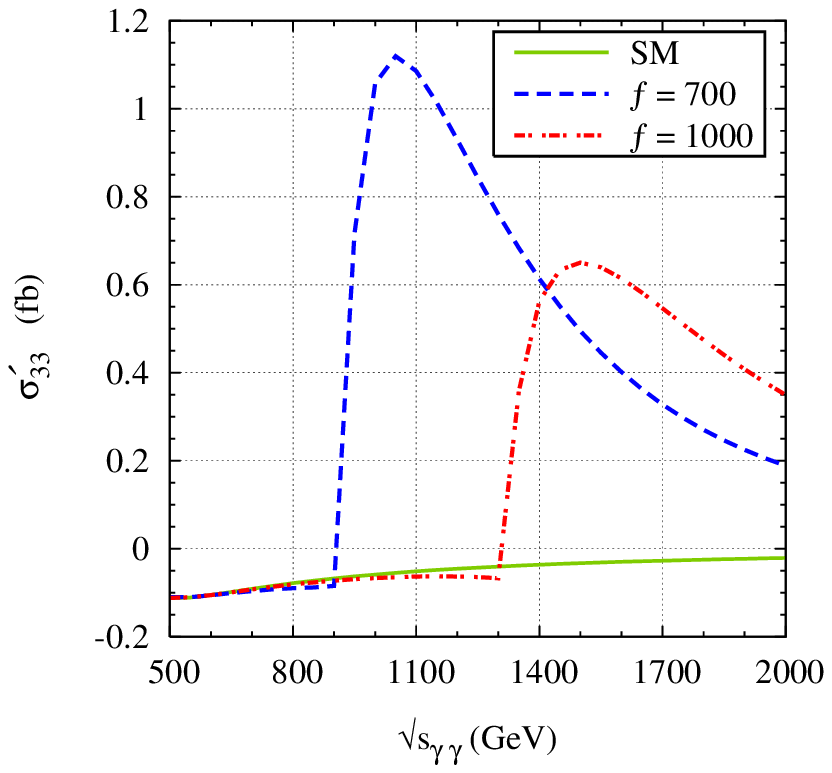,width=.6\textwidth} \hskip -3 cm
\epsfig{file=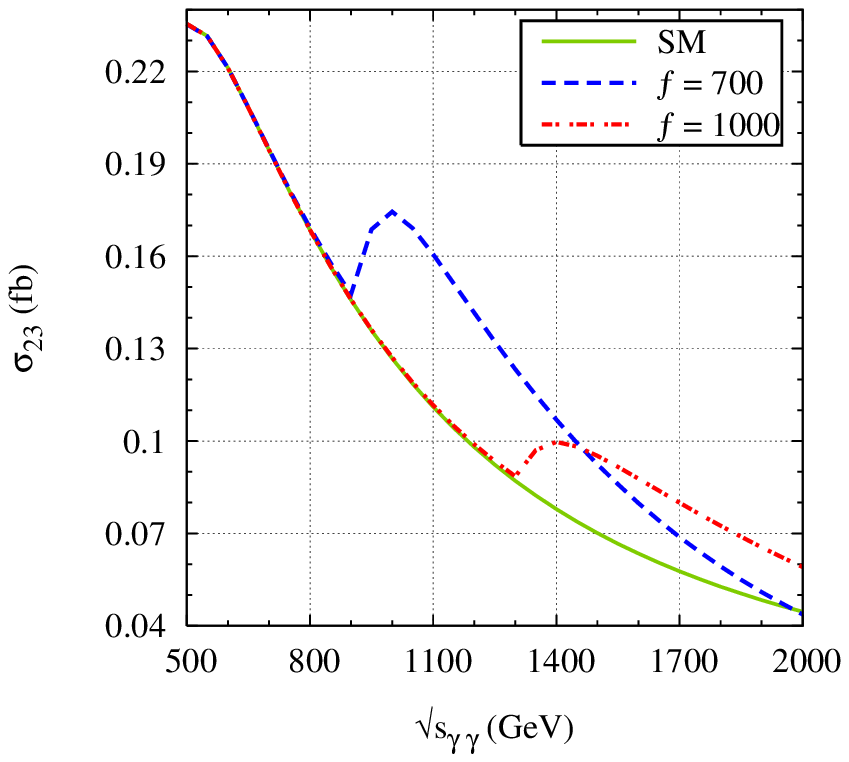,width=.6\textwidth}
\vskip -.5cm
\caption{\it Results for the cross-sections integrated in the range
$30 \le \theta^* \le 150$ for various values of $f$ (in GeV) in the LH
model with T-parity.}  
\label{fig:lh_intcross_2}
\end{figure}

 We should note, at this point, that the $\gggg$ scattering
proceeds through loops, both in the SM and in the LH models. In these
loops intermediate particles are pair produced (which is why LH models
with T-parity are particularly interesting as precision and
cosmological constraints on LH particle masses is much weaker
\cite{Asano:2006nr,Hubisz:2005tx}). In the SM these are dominated by
$W$ loops, leading to a peak in the SM cross-sections around the 
threshold of the $W$ pair production
\cite{Gounaris:1999gh}. Similarly, in the LH model (with and without
T-parity), the dominant contribution will come from the new heavy
$W$-boson and the Higgs particles (especially those that are doubly
charged), once we exceed the threshold for the pair production of
these particles. As such, we have plotted the various cross-sections
for a range of energies ($\sqrt{s_{\gamma\gamma}}$) well above the
threshold for the SM $W$-bosons, but in the vicinity of the pair
production energy for the new particles in the LH models, see figures 
(\ref{fig:lh_intcross_1},\ref{fig:lh_intcross_2}). Note further, that
we have integrated our differential cross-sections in the angular
range $30^\circ \le \theta^* \le 150^\circ$.   


\par We have plotted the SM value and LH value of the various
cross-sections introduced in the previous section in figures
(\ref{fig:lh_intcross_1},\ref{fig:lh_intcross_2}). As expected the 
deviation in the SM value of the cross-sections becomes visible around
the threshold of the pair production of LH particles. At present there
are very stringent constraints on the masses of LH particles in models
without T-parity \cite{Yue:2004xt,Han:2003wu,Buras:2004kq}, as
can be seen from figure (\ref{fig:lh_intcross_1}), the deviation from
SM values occurs at a very high value of
$\sqrt{s_{\gamma\gamma}}$. However, as has noted earlier, in LH models
with T-parity a comparatively lower value of the LH 
particle masses is allowed, which is reflected in the plots in figure
(\ref{fig:lh_intcross_2}).  

\par For $\gggg$ scattering the LH particles which contribute are the
charged gauge bosons ($W_H$), charged Higgs and charged fermions. The
present constraints on the LH models without T-parity forces the
masses of all the new heavy particles to be of the order of TeV. As we
are only concerned with charged particles the only parameters of
interest in the LH model without T-parity we will be $f$, $x_L$, $s$ (as
defined earlier in section 2).  The plots for LH model with T-parity
are shown in figure 
(\ref{fig:lh_intcross_2}), where we have chosen $\lambda_1/\lambda_2$
to be 1.   

\par In all cases we can get substantial deviations in the
cross-sections due to LH effects, these effects being prominent for
relatively lower values of $\sqrt{s_{\gamma\gamma}}$ for the models
with T-parity, where we have weaker constraints on the model
parameters. It should be noted that the $\sigma_3$ and $\sigma'_{33}$
provide the most interesting results, where the $\sigma_3$ is the only
cross-section with pronounced {\it ``dips"} (these being more
pronounced when T-parity is included in the model). The location of
these {\it ``dips"} being dependent on the model
parameters. The other feature of note in these plots are the
pronounced peaks in the $\sigma'_{33}$ cross-section. 
The LH model effects are more pronounced in $\sigma_3$ and
$\sigma'_{33}$. The SM values of the cross-sections $\sigma_3$ and
$\sigma'_{33}$ are relatively small as compared to the other
cross-sections, however, the new physics (the LH model here) effects
in these two cross-sections are very striking. These effects mainly
depend upon the LH parameter $f$ (the symmetry breaking scale of the
global symmetry). In LH models 
without T-parity the allowed value of $f$ is high, hence the masses of
the new heavy particles are high. This results in the deviations, in
LH results from SM results, as manifesting at higher 
values of the invariant mass. Whereas, in the case of T-parity models
a much lower value of $f$ is allowed. This now results in lower mass
values of T-odd 
particles; resulting in the onset of LH deviations at a much lower
invariant mass.

\par The results which we have presented for the process $\gggg$ are
rather generic and can be used as a probe for heavy charged gauge
bosons and charged scalars. In our results we have tried to focus
ourselves to the range of cm energy ($\sqrt{s_{\gamma\gamma}}$) which
is close to the threshold of the pair production of the particles.  
The deviations from SM results as shown in
figures (\ref{fig:lh_intcross_1},\ref{fig:lh_intcross_2}) will not be
observable in the proposed International Linear Collider (ILC), but
will be easily probed in a multi-TeV $e^+ e^-$ Compact linear collider
(CLIC); where it is proposed to build an $e^+ e^-$ linear collider
with a center of mass energy from 0.5 - 3TeV. Generically such a mode
should lead to $\gamma \gamma$ collisions at cm energies
$E^{\gamma\gamma}_{cm} \le 0.8 E^{ee}_{cm}$. Furthermore, 
the polarized cross-sections $\sigma_3$ and $\sigma'_{33}$ can be used
to test the spin structure of the particle loops which are responsible
for the $\gggg$ process \cite{Gounaris:1999gh}. In summary the $\gggg$
process is a very clean process which shall provide a very useful tool
for testing LH type models.  

%
%

%
%

\begin{acknowledgments}
The work of SRC, NG and AKG was supported by the Department of Science
\& Technology (DST), India under grant no SP/S2/K-20/99. The work of
A.S.C. was supported by the Japan Society for the Promotion of Science
(JSPS), under fellowship no P04764. The work of N.G. was partly
supported by JSPS grant no P06043. NG would like to thank Yukawa
Institute of Theoretical Physics (YITP) for local hospitality where
this work was initiated. 
\end{acknowledgments}

\appendix

%
%

\section{Helicity amplitudes \label{appen:a}}\cleqn

\par As noted earlier in this paper, for the process;
\beq
\gamma(p_1, \lambda_1) \gamma(p_2, \lambda_2) \to \gamma(p_3,
\lambda_3) \gamma(p_4, \lambda_4) ,   
\eeq
the helicity amplitudes can be denoted $F_{\lambda_1 \lambda_2
\lambda_3 \lambda_4}(\hat{s}, \hat{t}, \hat{u})$, where the momenta
and helicities of the incoming and outgoing photons are as denoted in
the above equation, and where we have used the Mandelstam variables $\hat{s}
= (p_1 + p_2)^2$, $\hat{t} = (p_1 - p_3)^2$ and $\hat{u} = (p_1 - p_4)^2$. Recall
that the use of Bose statistics, crossing symmetry, and parity and
time inversion invariance results in the 16 possible helicity
amplitudes as being expressible in terms of just three
amplitudes. Namely, $F_{+++-}(\hat{s}, \hat{t}, \hat{u})$,
$F_{++--}(\hat{s}, \hat{t}, \hat{u})$ and $F_{++++}(\hat{s}, \hat{t}, \hat{u})$ through 
\cite{Jikia:1993tc}. 
\bea
F_{\pm \pm \mp \pm} (\hat{s}, \hat{t}, \hat{u}) & = & F_{\pm \mp \pm
\pm}(\hat{s}, \hat{t}, \hat{u}) = F_{\pm \mp \mp \mp}(\hat{s},
\hat{t}, \hat{u}) = F_{---+}(\hat{s}, \hat{t}, \hat{u}) \nonumber \\  
& = & F_{+++-}(\hat{s}, \hat{t}, \hat{u}) , \\
F_{--++}(\hat{s}, \hat{t}, \hat{u}) & = & F_{++--}(\hat{s}, \hat{t},
\hat{u}) , \\  
F_{\pm \mp \pm \mp}(\hat{s}, \hat{t}, \hat{u}) & = & F_{----}(\hat{u},
\hat{t}, \hat{s}) = F_{++++}(\hat{u}, \hat{t}, \hat{s}) , \\  
F_{\pm \mp \mp \pm}(\hat{s}, \hat{t}, \hat{u}) & = & F_{\pm \mp \pm
\mp}(\hat{s}, \hat{u}, \hat{t}) = F_{++++}(t, s, u) =
F_{++++}(\hat{t}, \hat{u}, \hat{s}) .  
\eea

\par Note that in expressing the SM and LH helicity amplitudes we
shall use the notation of B, C and D functions as given in
\cite{Hagiwara:1994pw}. 
The $B_0$, $C_0$ and $D_0$ are the usual one-loop functions
first introduced by Passarino and Veltman \cite{Passarino:1978jh}. 
The charged gauge boson contributions to the helicity amplitudes can
be written as \cite{Jikia:1993tc}:  
\bea
\ds \frac{F_{++++}^W (\hat{s}, \hat{t}, \hat{u})}{\alpha^2} & = & 12 -
12 \left( 1 + \ds \frac{2 \hat{u}}{\hat{s}} \right) B_0 (\hat{u}) - 12
\left( 1 + \frac{2 \hat{t}}{\hat{s}} \right) B_0 (\hat{t}) \nonumber
\\  
&& \ds \hspace{0.2cm} + \frac{24 m_W^2 \hat{t} \hat{u}}{\hat{s}} D_0
(\hat{u}, \hat{t}) + 16 \left( 1 - \frac{3 m_W^2}{2 \hat{s}} - \frac{3
\hat{t} \hat{u}}{4 \hat{s}^2} \right) \nonumber \\  
&& \times \ds \left[ 2 \hat{t} C_0 (\hat{t}) + 2 \hat{u} C_0 (\hat{u})
- \hat{t} \hat{u} D_0 (\hat{t}, \hat{u}) \right] + 8 \left( \hat{s} -
m_W^2 \right) \left( \hat{s} - 3 m_W^2 \right) \nonumber \\  
&& \times \ds \left[ D_0 (\hat{s}, \hat{t}) + D_0 (\hat{s}, \hat{u}) +
D_0 (\hat{t}, \hat{u}) \right] , \label{amp1} \\  
\ds \frac{F_{+++-}^W (\hat{s}, \hat{t}, \hat{u})}{\alpha^2} & = & - 12
+ 24 m_W^2 \left[ D_0 (\hat{s}, \hat{t}) + D_0 (\hat{s}, \hat{u}) +
D_0 (\hat{t}, \hat{u}) \right] \nonumber \\  
&& \hspace{0.2cm} + 12 m_W^2 \hat{s} \hat{t} \hat{u} \left[ \ds
\frac{D_0 (\hat{s}, \hat{t})}{\hat{u}^2} + \frac{D_0 (\hat{s},
\hat{u})}{\hat{t}^2} + \frac{D_0 (\hat{t}, \hat{u})}{\hat{s}^2} 
\right] \nonumber \\ 
&& \hspace{0.2cm} - 24 m_W^2 \left( \ds \frac{1}{\hat{s}} +
\frac{1}{\hat{t}} + \frac{1}{\hat{u}} \right) \left[ \hat{t} C_0
(\hat{t}) + \hat{u} C_0 (\hat{u}) + \hat{s} C_0 (\hat{s}) \right] , 
\label{amp2} \\  
\ds \frac{F_{++--}^W (\hat{s}, \hat{t}, \hat{u})}{\alpha^2} & = & - 12
+ 24 m_W^2 \left[ D_0 (\hat{s}, \hat{t}) + D_0 (\hat{s}, \hat{u}) +
D_0 (\hat{t}, \hat{u}) \right] . \label{amp3}  
\eea
The contributions from a fermion of charge $Q_f$ and mass $m_f$ to the
helicity amplitudes can then be written as \cite{Jikia:1993tc}: 
\bea
\ds \frac{F_{++++}^f (\hat{s}, \hat{t}, \hat{u})}{\alpha^2 Q_f^4} & =
& - 8 + 8 \left( \ds 1 + \frac{2 \hat{u}}{\hat{s}} \right) B_0
(\hat{u}) + 8 \left( 1 + \frac{2 \hat{t}}{\hat{s}} \right) B_0
(\hat{t}) \nonumber \\  
&& \hspace{0.2cm} \ds - 8 \left( \frac{\hat{t}^2 +
\hat{u}^2}{\hat{s}^2} - \frac{4 m_f^2}{\hat{s}} \right) \left[ \hat{t}
C_0 (\hat{t}) + \hat{u} C_0 (\hat{u}) \right] \nonumber \\  
&& \hspace{0.2cm} \ds + 8 m_f^2 \left( \hat{s} - 2 m_f^2 \right)
\left[ D_0 (\hat{s}, \hat{t}) + D_0 (\hat{s}, \hat{u}) \right]
\nonumber \\  
&& \hspace{0.2cm} - 4 \left[ 4 m_f^4 - \left( 2 \hat{s} m_f^2 +
\hat{t} \hat{u} \right) \frac{\hat{t}^2 + \hat{u}^2}{\hat{s}^2} +
\frac{4 m_f^2 \hat{t} \hat{u}}{\hat{s}} \right] D_0 (\hat{t}, \hat{u})
, \label{amp4} \\  
\ds F_{+++-}^f (\hat{s}, \hat{t}, \hat{u}) & = & \ds - \frac{2}{3}
Q_f^4 \left\{ F_{+++-}^W (\hat{s}, \hat{t}, \hat{u}) ; m_W \to m_f
\right\} , \label{amp5} \\  
\ds F_{++--}^f (\hat{s}, \hat{t}, \hat{u}) & = & \ds - \frac{2}{3}
Q_f^4 \left\{ F_{++--}^W (\hat{s}, \hat{t}, \hat{u}) ; m_W \to m_f
\right\} . \label{amp6}   
\eea
\noindent As discussed earlier, the LH model introduces several new
particles, including new scalar particles. As such the contribution
from new scalar particles of mass $m_s$ and charge $Q_s$ to the
helicity amplitudes can be written as \cite{Jikia:1993tc}: 
\bea
\ds \frac{F_{++++}^s (\hat{s}, \hat{t}, \hat{u})}{\alpha^2 Q_s^4} & =
& 4 - 4 \left( \ds 1 + \frac{2 \hat{u}}{\hat{s}} \right) B_0 (\hat{u})
- 4 \left( 1 + \frac{2 \hat{t}}{\hat{s}} \right) B_0 (\hat{t})
\nonumber \\  
&& \hspace{0.2cm} + \ds \frac{8 m_s^2 \hat{t} \hat{u}}{\hat{s}} D_0 (
\hat{t}, \hat{u} ) - \frac{8 m_s^2}{\hat{s}} \left( 1 + \frac{\hat{u}
\hat{t}}{2 m_s^2 \hat{s}} \right) \nonumber \\  
&& \times \ds \left[ 2 \hat{t} C_0 (\hat{t}) + 2 \hat{u} C_0 (\hat{u})
- \hat{t} \hat{u} D_0 (\hat{t}, \hat{u}) \right] + 8 m_s^4 \left[ D_0
(\hat{s}, \hat{t}) \right. \nonumber \\  
&& \left. \hspace{0.5cm} \ds + D_0 (\hat{s}, \hat{u}) + D_0 (\hat{t},
\hat{u}) \right] , \label{amp7} \\  
F_{+++-}^s (\hat{s}, \hat{t}, \hat{u}) & = & \ds \frac{1}{3} Q_s^4
\left\{ F_{+++-}^W (\hat{s}, \hat{t}, \hat{u}) ; m_W \to m_s \right\}
, \label{amp8} \\  
F_{++--}^s (\hat{s}, \hat{t}, \hat{u}) & = & \ds \frac{1}{3} Q_s^4
\left\{ F_{++--}^W (\hat{s}, \hat{t}, \hat{u}) ; m_W \to m_s \right\}
. \label{amp9}  
\eea
\noindent Whilst, new fermions and bosons shall be incorporated with
helicity amplitudes presented in equations (\ref{amp1}-\ref{amp6}). 

%
%

\section{Input parameters}\cleqn

$$m_H = 120 \ {\rm GeV}~,~~ 
v = 246 \ {\rm GeV}~,~~ \alpha = \frac{1}{130}
~~,~~ g^2 = 0.34 ~~,~~ g'^2 = 0.12 $$

%
%

\end{document}